\documentclass[preprint,amsmath,amssymb,aps]{revtex4-1}
\usepackage{graphicx}
\usepackage{dcolumn}
\usepackage{booktabs}
\usepackage{bm}
\usepackage{subfig}
\usepackage{amsmath}

\usepackage{color}

\newcommand{\degree}{^\circ}
\graphicspath{{FIG/}}

\begin{document}

\preprint{APS/123-QED}

\title{Spontaneous shrinkage of droplet on  wetting surface in phase-field model}

\author{Chunhua Zhang}
\affiliation{
State Key Laboratory of Coal Combustion, Huazhong University of Science and Technology, Wuhan 430074, China}
\author{Zhaoli Guo}
\email{zlguo@hust.edu.cn}
\affiliation{
State Key Laboratory of Coal Combustion, Huazhong University of Science and Technology, Wuhan 430074, China}
\date{\today}

\begin{abstract}
Phase field theory is widely used to model multi-phase flows.  A drop can shrink or grow spontaneously due to the redistribution of interface and bulk energies to minimize the system energy. In this paper, the spontaneous behaviour of a drop on a flat surface is investigated. It is found that there exists a critical radius dependent on the contact angle, the domain size and the interface width, below which the droplet will eventually disappear. In particular, the critical radius can be very large when the contact angle is hydrophilic.
The theoretical prediction of the critical radius is verified numerically by simulating a drop on a surface with various contact angles, the domain sizes and the interface widths.
\end{abstract}
\maketitle

In recent years, the phase-field model has received extensive attention in the fields of natural sciences and engineering applications, such as solidification and  crystal growth~\cite{karma2001phase,rojas2015phase,nestler2005multicomponent},
 crack propagation~\cite{ambati2015review,pons2010helical}, multiphase flows~\cite{badalassi2003computation,shen2015decoupled} and moving contact line~\cite{ding2007wetting,yue2011can}. Within this framework, the interface between two-phase fluids is treated as a thin layer of several grid sizes instead of a sharp interface. The interfacial profile  can be identified by a suitably defined phase field variable, which is usually described by certain diffusive models such as the Cahn-Hilliard equation or the Allen-Cahn equation. A striking advantage of the phase-field model is free of  explicitly tracking the interface between two-phase fluids, which leads to the phase field method easy to implement for interfacial dynamics in complex flows~\cite{badalassi2003computation,teigen2011diffuse}.

However, it is found that the size of a circular bubble can shrink to zero due to the dissipative mechanism in the phase field theory framework~\cite{liu2003phase}.
Yue~\emph{et al.}~\cite{yue2004diffuse} noticed that
the interface can shrink slightly even  when the interface of a drop is initialized by the exact hypertangential profile.
Similarly, Lee~\emph{et al.}~\cite{lee2006eliminating} observed that the radius of a droplet decreases first and then reaches a steady-state value in simulations.
To understand this behavior, Yue \emph{et al.}~\cite{yue2007spontaneous}
made a theoretical analysis of the Cahn-Hilliard model and found there existed a critical drop radius dependent on the  domain size and interface width,  below which the droplet will eventually disappear.
Following the same procedure of Yue~\emph{et al}, zheng \emph{et al.}~\cite{zheng2014shrinkage} demonstrated that there is a similar critical radius for a pseudo-Vander Waals fluid in the single-fluid diffuse interface method. However,
in both works, no solid walls are involved and the result is inapplicable to systems with solid bodies.
In deed, it was found numerically that a drop on a solid surface can shrink with different rates for varying wetting properties~\cite{zheng2014shrinkage}.
As  the wetting of liquid on the solid surface is of great importance in defining the flow characteristics~\cite{bonn2009wetting,sohrabi2011safe}, it is important to
study the critical behavior of a droplet on a solid surface. In this paper we will present an such analysis in view of the Cahn-Hillard equation.

\begin{figure}[htp]
\centering
\subfloat{\includegraphics[width=0.5\textwidth]{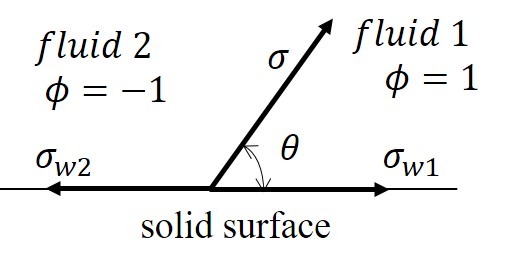}}
\caption{Schematic of a static contact angle.}
\label{fig:sketchContactLine}
\end{figure}

The Cahn-Hillard equation for an immiscible two-phase system can be written as
\begin{equation}\label{eq:CH}
\frac{\partial \phi}{\partial t}=\nabla\cdot(M\nabla\mu),
\end{equation}
where $\phi$ is the phase-field variable, i.e., $\phi_1=1$ in fluid $1$ and $\phi_2=-1$ in fluid $2$, $M$ is the mobility which is assumed to be constant in the present study and  $\mu$ is the chemical potential defined from the mixing energy in the whole domain. For a two-phase system with a solid surface,
The total mixing energy combines a bulk contribution from the Ginzburg-Landau free energy and a surface energy contribution from the substrate~\cite{cahn1958free},
\begin{equation}\label{eq:mixng}
F=\int_{\Omega}\left[f(\phi)+\frac{\kappa}{2}|\nabla \phi|^2\right]d \Omega+\int_{\partial\Omega}f_w(\phi)d \partial\Omega,
\end{equation}
where $f(\phi)=\beta(\phi^2-1)^2$  is a double well potential, $\kappa$ is a constant and $\partial\Omega$ is the boundary of computational domain $\Omega$. Both $\beta$ and $\kappa$ are related to  the interface tension $\sigma$ and the  interface thickness $W$, i.e., $\beta=3\sigma/(4W)$, $\kappa=3/(8\sigma W)$.  The wall energy $f_w$ can be given by~\cite{yue2010sharp},
\begin{equation}\label{eq:wallEnergy}
f_w(\phi)=-\sigma \cos\theta\frac{\phi(3-\phi^2)}{4}+\frac{\sigma_{w1}+\sigma_{w2}}{2},
\end{equation}
where $\sigma_{w1}$ ($\sigma_{w2}$) is the fluid-solid interfacial tension between the wall and fluid 1 (fluid 2), $\theta$ is the static contact angle.  As shown in Fig.~\ref{fig:sketchContactLine}, the contact angle satisfies the Young's equation, i.e., $\sigma_{w2}-\sigma_{w1}=\sigma\cos\theta$,.
The chemical potential $\mu$ is defined by
\begin{equation}\label{eq:chemical_potential_energy}
\mu=f'(\phi)-\kappa \Delta \phi.
\end{equation}

For the Cahn-Hilliard equation, it can be easily identified that  the total mass is conserved,
\begin{equation}\label{eq:conserved}
\frac{d}{dt}\int_{\Omega} \phi d\Omega=\int_{\partial\Omega}M\nabla \mu\cdot \bm n  d\Omega=0,
\end{equation}
and the total energy of the system is non-increasing over time,
\begin{equation}\label{eq:energydecrease}
\begin{aligned}
\frac{d F(\phi)}{dt}&=-\int_{\Omega}M|\nabla\mu|^2 d\Omega,
\end{aligned}
\end{equation}
with the following boundary conditions,
\begin{align}\label{eq:boundary}
  \bm u_{\partial \Omega}&=0,\\
  \bm n\cdot \nabla\mu|_{\partial\Omega}&=0,\\
(\kappa\bm n\cdot \nabla \phi+f_w'(\phi))|_{\partial\Omega}&=0,
\end{align}
where $\bm n$ is the unit normal vector pointing into the wall. These two properties play important roles in analyzing the critical radius of a drop on a flat solid surface.

\begin{figure}[htp]
\centering
\subfloat{\includegraphics[width=0.75\textwidth,trim={3.8cm 14cm 8cm 8cm},clip]{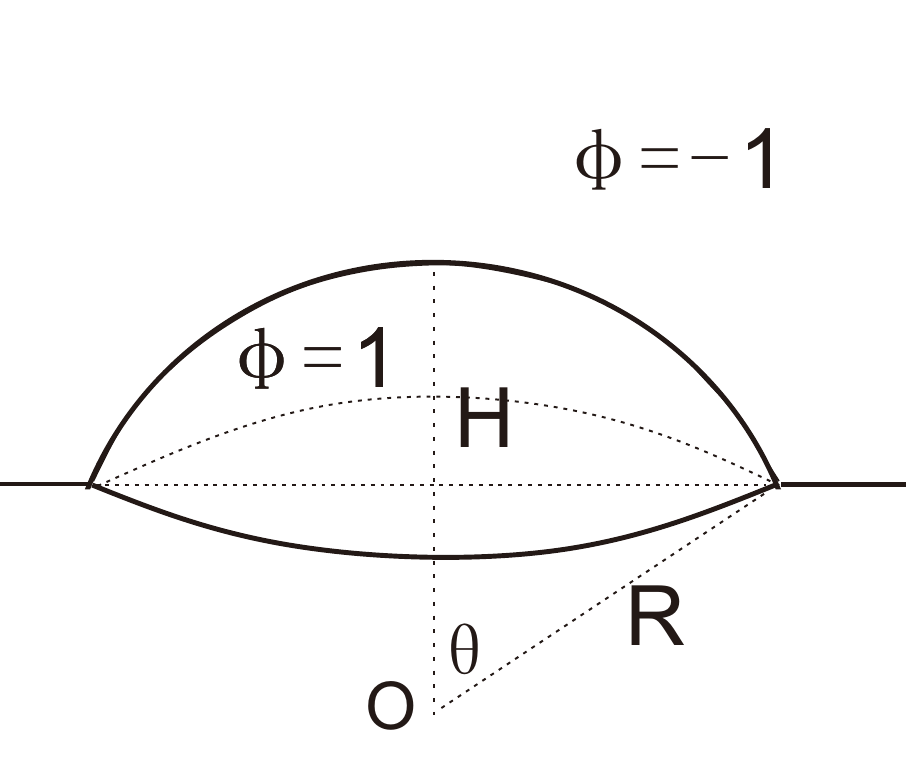}}
\caption{Schematic of a drop on a flat solid surface.}
\label{fig:half_solid_le90}
\end{figure}

We consider a  general case where a drop with initial radius $R_0$ is in contact with a flat solid surface in a domain with size $L\times L\times L$, as shown in Fig.~\ref{fig:half_solid_le90}.
Following the same assumption of Ref.\cite{yue2007spontaneous}, there is an equal and uniform shift $\delta \phi$ inside and outside of the drop when the drop radius shrink to $R$ from $R_0$. In addition, we further assume that the contact angle remains unchanged and the shape of the drop maintains a spherical cap during shrinking.  With these assumptions, we can estimate the critical radius of the drop on the solid surface.

\begin{figure}[htp]
\centering
\subfloat{\includegraphics[width=0.75\textwidth]{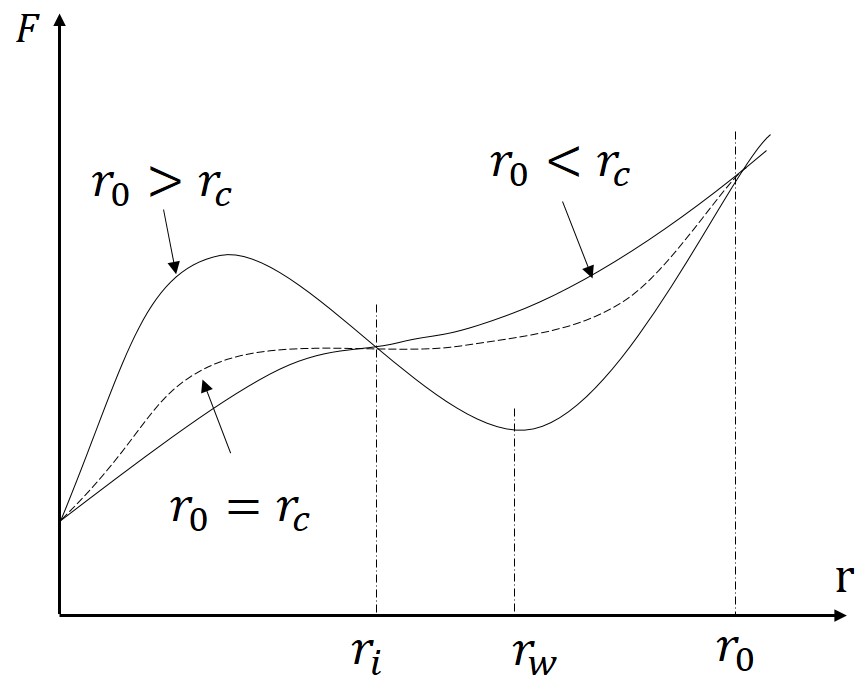}}
\caption{Schematic of the mixing energy $F$ as a function of $r$.}
\label{fig:mixing_energy}
\end{figure}

\begin{figure}[htp]
\centering
\subfloat[]{\includegraphics[width=0.75\textwidth]{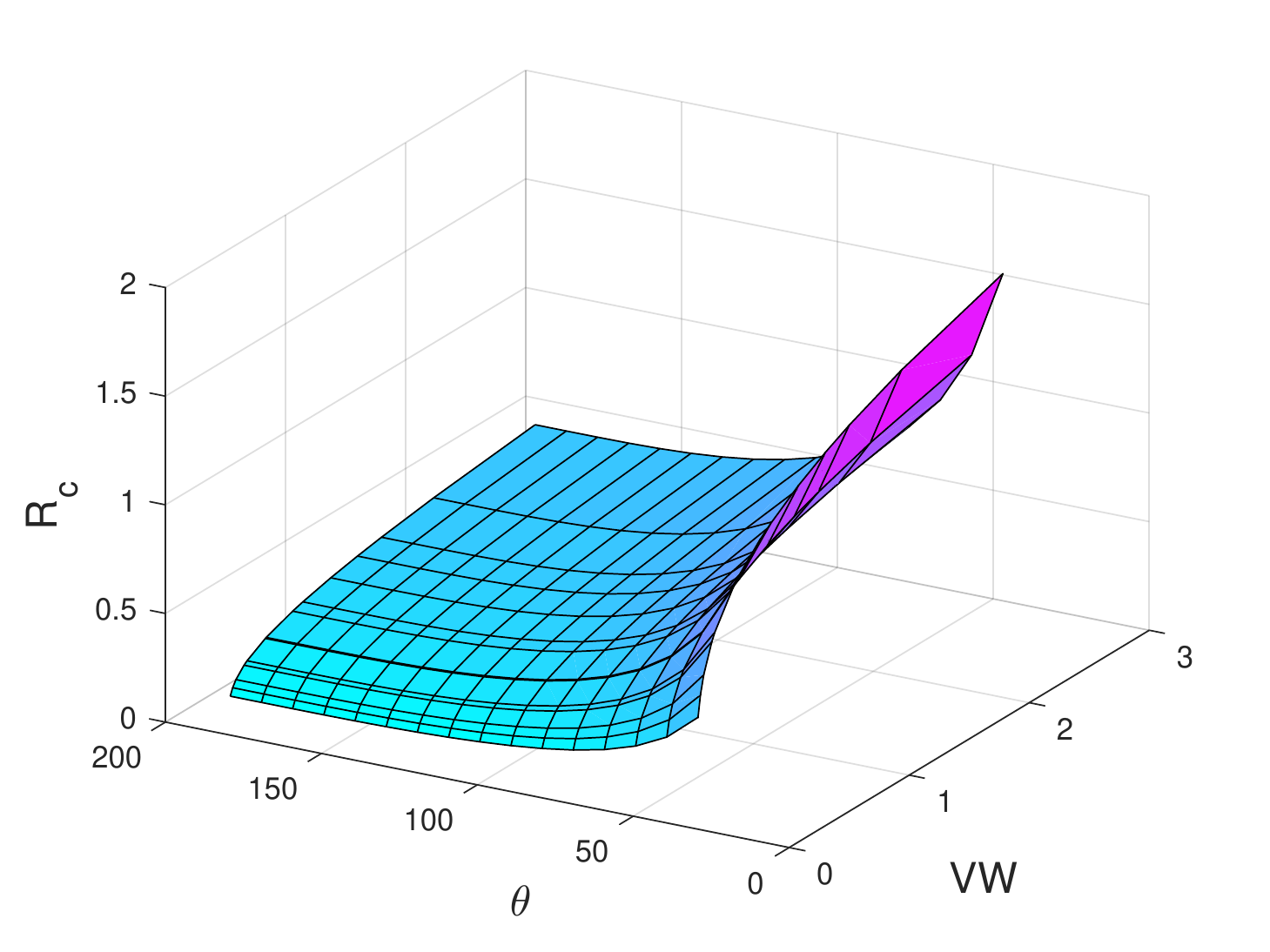}}
\caption{The critical drop  radius $R_c$ as a function of $VW$ and contact angles  based on Eq.~(\ref{eq:critical3D}).}
\label{fig:3D_theoritical}
\end{figure}

From Fig.~\ref{fig:half_solid_le90}, it can be seen that the volume of spherical cap is $V_1=\pi(3R-H)H^2/3$ with $H=(1-\cos\theta)R$ being the height of the spherical cap. According to the conservation property of the phase field variable, the  shift of $\phi$ is
\begin{equation}\label{eq1:shrinks1}
\delta \phi=\frac{2\pi (R_0^3-R^3)\theta_c}{3V},
\end{equation}
where $\theta_c=(\cos\theta)^3-3\cos\theta+2$ and $V=L^3$ is the volume of the domain.
Although $R_0^3-R^3$ is finite here, it can be confirmed that $\delta \phi\sim r_c^3/V\sim W^{3/4}/V^{1/4}\ll 1$.
By neglecting cubic and quartic terms in $\delta \phi$, the system free energy $F(R)$ can be approximated by

\begin{equation}\label{eq2:2D_totalEnergy}
\begin{split}
F(R)&=\sigma S+\sigma_{w1} d_1+\sigma_{w2} (D-d_1)
+\beta  (\phi_1^2-1)^2 V_1 +\beta (\phi_2^2-1)^2 (V-V_1)\\
&\simeq \sigma S+(\sigma_{w1}-\sigma_{w2}) d_1+\sigma_{w2} D
+4\beta V \delta \phi^2 ,
\end{split}
\end{equation}
where  $S=2\pi RH$ is the area of the surface of spherical cap, $d_1=\pi(\sin\theta R)^2$ is the area between  fluid $1$ and the solid surface and $D$ is the total interface area between  the fluids and the solid surface.
As the surface energy of the drop in the three-dimensional (3D) space is proportional to area (for two-dimensional case, the surface energy is proportional to perimeter), the above equation can be rewritten as,
\begin{equation}\label{eq:F_r}
F(r)=2\sigma\pi r(1-\cos\theta)-\sigma \pi r\cos\theta \sin^2\theta +\sigma_{w2} D
+\frac{16\beta \pi^2 (R_0^2+r^3-2r^{3/2}R_0^3)\theta_c^2}{9V},
\end{equation}
where $r=R^2$. As shown in Fig.~\ref{fig:mixing_energy}, the critical radius of the drop should correspond to an inflection point of the system energy, below which the energy decreases continuously. Thus we explicitly give the first to third derivatives of $F(r)$ with respect to $r$,
\begin{equation}\label{eq:DF_r}
\frac{\partial F}{\partial r}=2\sigma\pi(1-\cos\theta)-\pi\sigma \cos\theta \sin^2\theta
+\frac{16\beta \pi^2 (3r^2-3\sqrt{r}R_0^3)\theta_c^2}{9V},
\end{equation}
\begin{equation}\label{eq:derivative2}
\frac{\partial^2 F}{\partial r^2}=
\frac{16\beta\pi^2\left(6r- 3R_0^3/2/\sqrt{r}\right)\theta^2}{9V},
\end{equation}
\begin{equation}\label{eq:derivative3}
\frac{\partial^3 F}{\partial r^3}=
\frac{16\beta \pi^2\left(6+\frac{3R_0^3}{4r^{3/2}}\right)\theta^2}{9V}.
\end{equation}
Based on Eq.~(\ref{eq:derivative3}),
a inflexion point $R_i$ can be determined from $\frac{\partial^2F}{\partial r^2}=0$, leading to $R_i=2^{2/3}R_0^2/4$. A vanishing drop requires
$\frac{\partial F}{\partial r}|_{R=R_i}=0$. As a result, the critical radius is given by
\begin{equation}\label{eq:critical3D}
R_c=\left(\frac{4}{\cos^3\theta-3\cos\theta +2}  \right)^{1/4} \left(\frac{2^{-4/3}VW}{3\pi}\right)^{1/4},
\end{equation}
suggesting that the drop with $R_0<R_c$ will disappear eventually due to energy minization.
The resulting formula is suitable to all contact angles (i.e., $\theta\in(0,\pi]$) although the contact angle in the above derivation is less than $\pi/2$.
In particular, as $\theta=\pi$,  the solid surface is completely non-wetting, the wall effect can be ignored, and the formula of the critical radius can be reduced to the result for a droplet without walls in~\cite{yue2007spontaneous}.
Similarly, the critical radius of a two-dimensional (2D) drop  can be determined as
 \begin{equation}\label{eq:criticalRadius2D}
R_c=\left(\frac{\pi}{\theta-\sin\theta\cos\theta}\right)^{1/3}\left(\frac{\sqrt{3}VW}{16\pi}\right)^{1/3}.
 \end{equation}
The above formula with $\phi=\pi$ is consistent with the one without walls in Ref.~\cite{yue2007spontaneous} as well.
The critical radius ($R_c$) profile of a drop with the products of the domain sizes and the interface widths $VW$ and contact angles $\theta$ based on Eq.~(\ref{eq:critical3D}) is shown in Fig~\ref{fig:3D_theoritical}.
It can be observed that the critical radius is largely determined by  the domain size and the contact angle.

\begin{figure}[htp]
\centering
\subfloat[]{\includegraphics[width=0.5\textwidth]{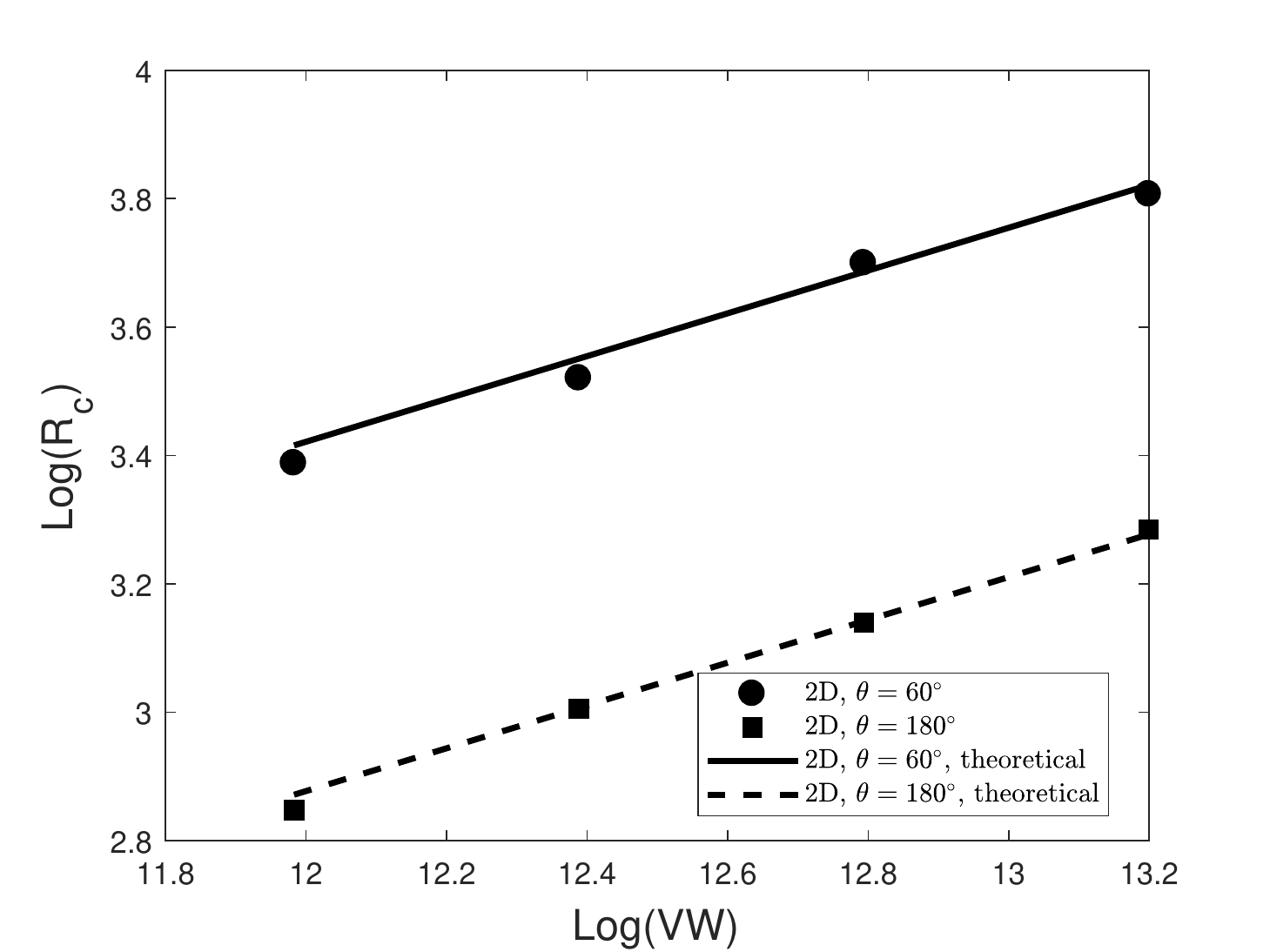}}~
\subfloat[]{\includegraphics[width=0.5\textwidth]{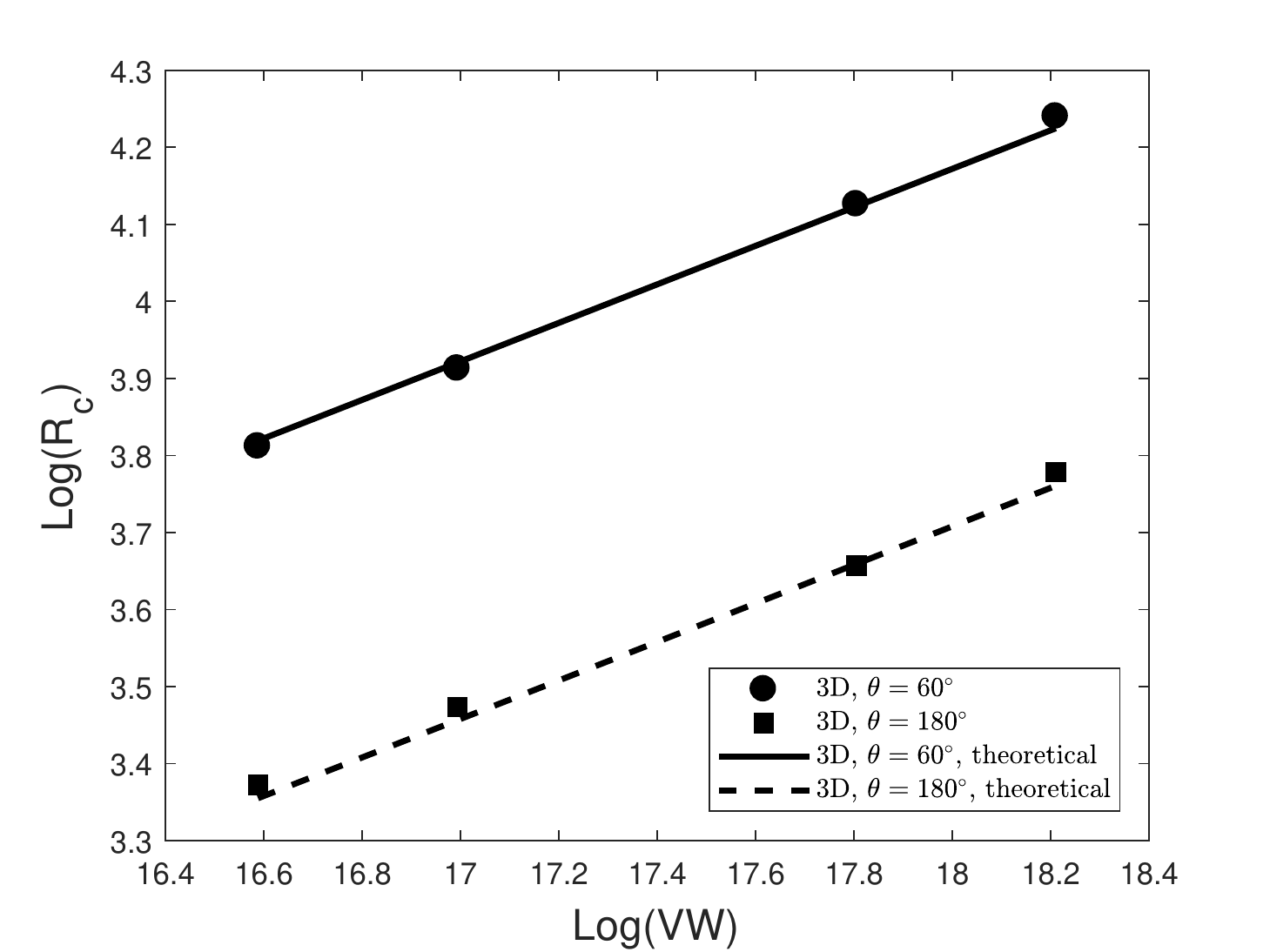}}
\caption{Comparison between theoretical and numerical results of the critical radius for (a) 2D drop and (b) 3D drop. The range of $VW$ is calculated by using $L=200, 300$ and $W=4, 6$.}
\label{fig:2D_radius_vs_VW}
\end{figure}

\begin{figure}[htp]
\centering
\subfloat{\includegraphics[width=0.75\textwidth]
{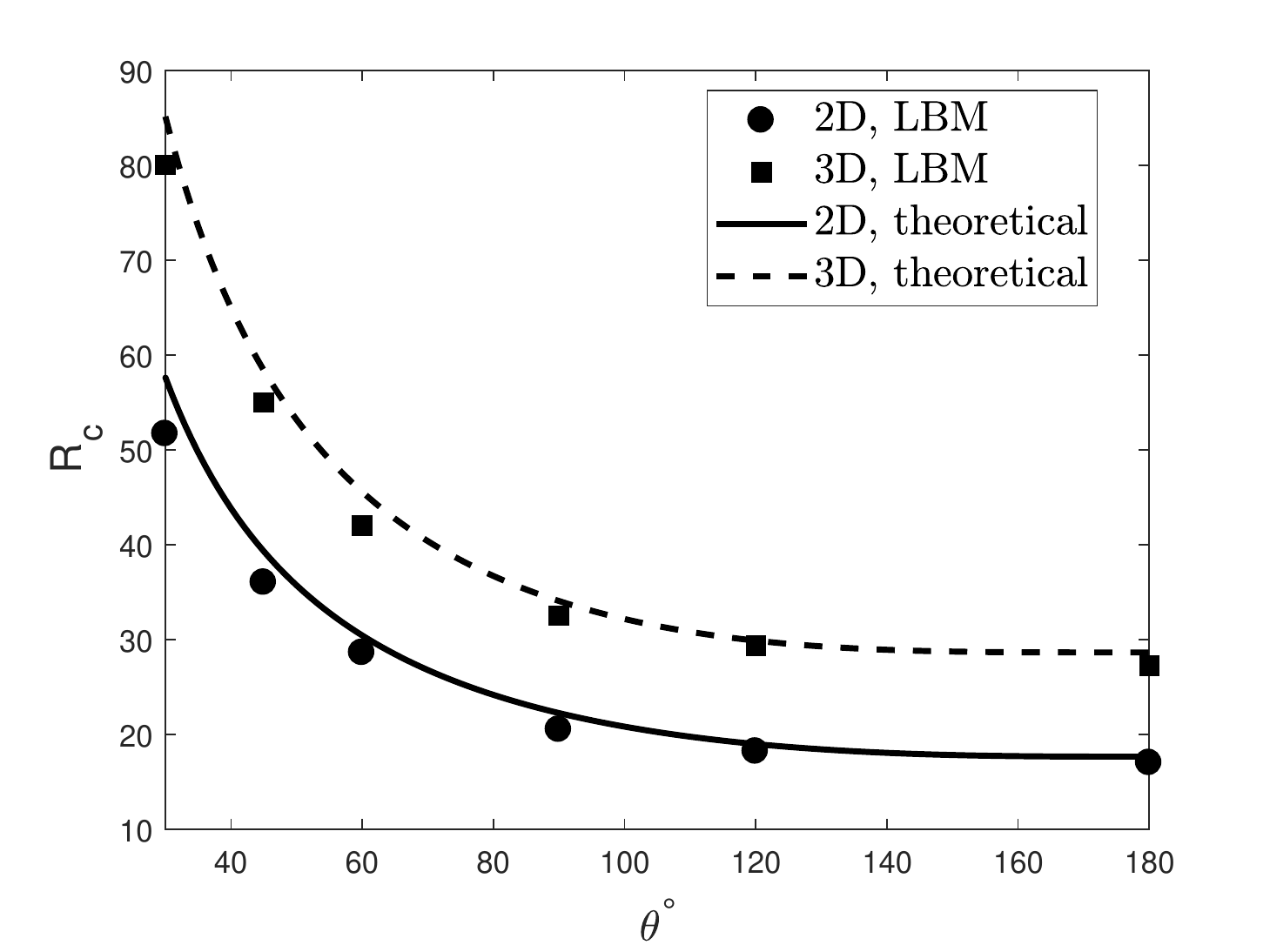}}
\caption{ The critical drop radius  with various contact angles $\theta$. The solid and dashed curves represent the theoretical predictions based on Eqs.~(\ref{eq:criticalRadius2D}) and (\ref{eq:critical3D}), respectively.}
\label{fig:2D_raidus_vs_angle}
\end{figure}

We now demonstrate numerically
the theoretical critical radius of a drop
on the solid surface with various contact angles varying from $\pi/6$ to $\pi$ using the lattice Boltzmann method~\cite{zhang2019high}. A uniform cartesian grid is used and all simulation parameters are expressed in lattice units.
First, we verify the relationship between the critical radius and the domain size as well as interface width.
In the simulations, the domain is $L\times L$ for 2D simulations and $L\times L\times L/2$ for $3D$ simulations. The parameters $\sigma$ is fixed at $0.05$.
Figure.~\ref{fig:2D_radius_vs_VW} shows the results as $V=200$ and $300$ with $W=4$ and $6$. It can be seen that the numerical results are consistent with the theoretical predictions. Then, we test the theoretical formulas for both 2D drops and 3D drops with various contact angles.
The results of 2D droplets are shown in Fig.~\ref{fig:2D_raidus_vs_angle}. Again, good agreements with the theoretical predictions are observed.

\begin{figure}[htp]
\centering
\subfloat[]{\includegraphics[width=0.5\textwidth]{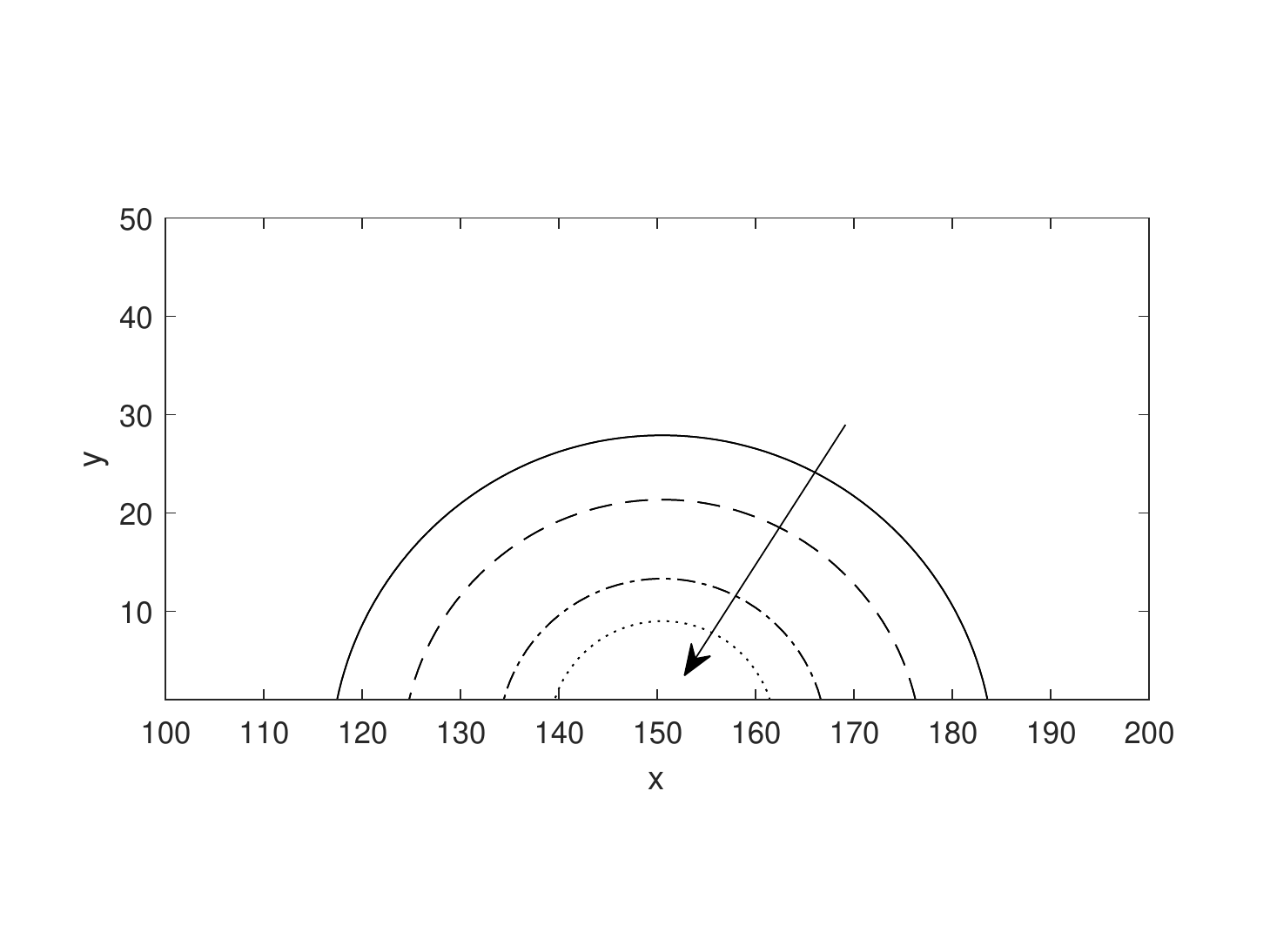}}
\subfloat[]{\includegraphics[width=0.5\textwidth]{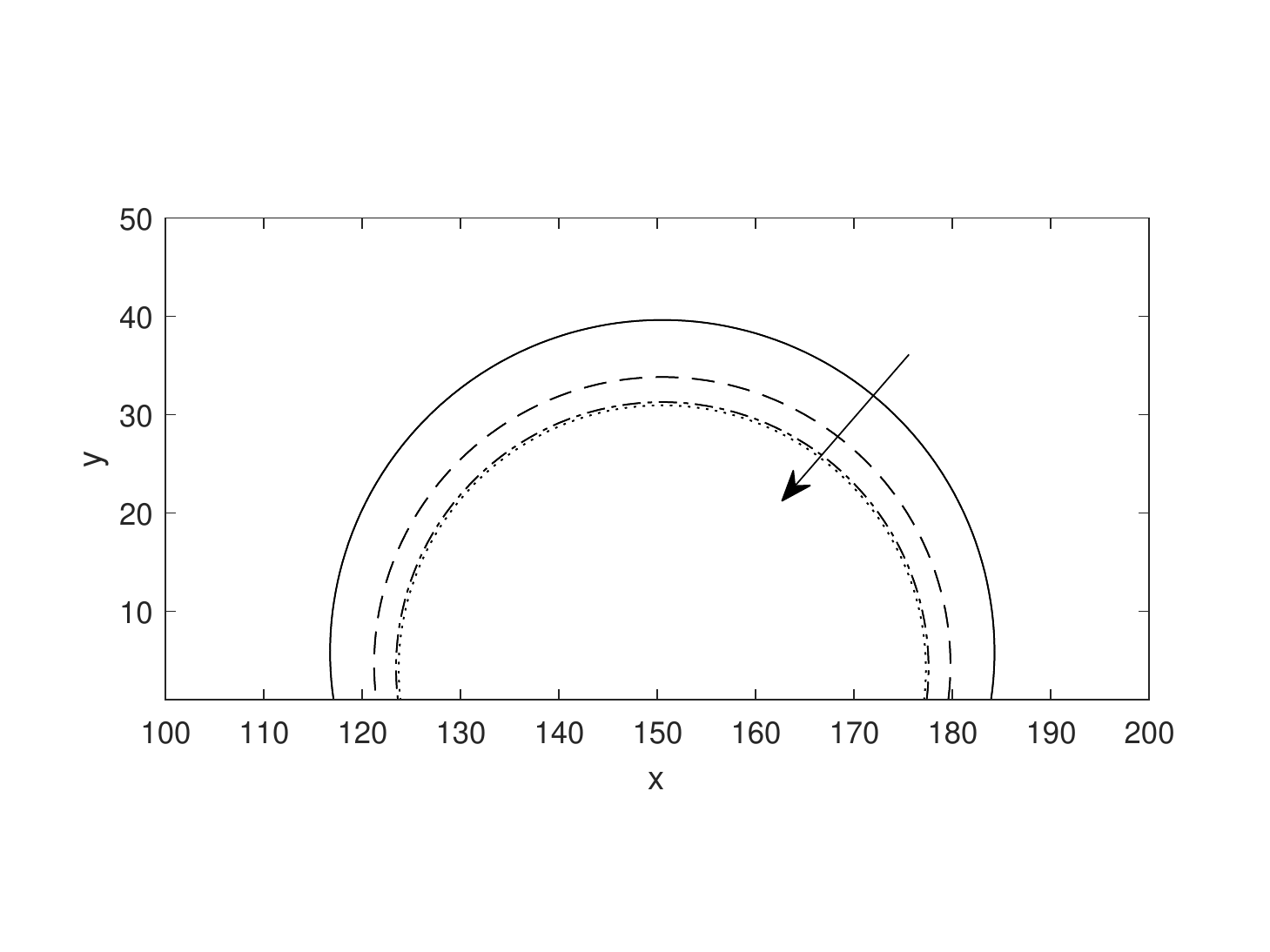}}
\caption{Time evolution of the interface contours  ($\phi=0$) of 2D drop with initial radius $R_0=33$ for (a) $\theta=80\degree (R_c=36)$ and (b)$\theta=100\degree (R_c=31)$. Arrows indicate the time $t/10^6=0, 5, 10, 15$.}
\label{fig:2Dtheta80_100}
\end{figure}

\begin{figure}[htp]
\centering
\subfloat[]{\includegraphics[width=0.75\textwidth]{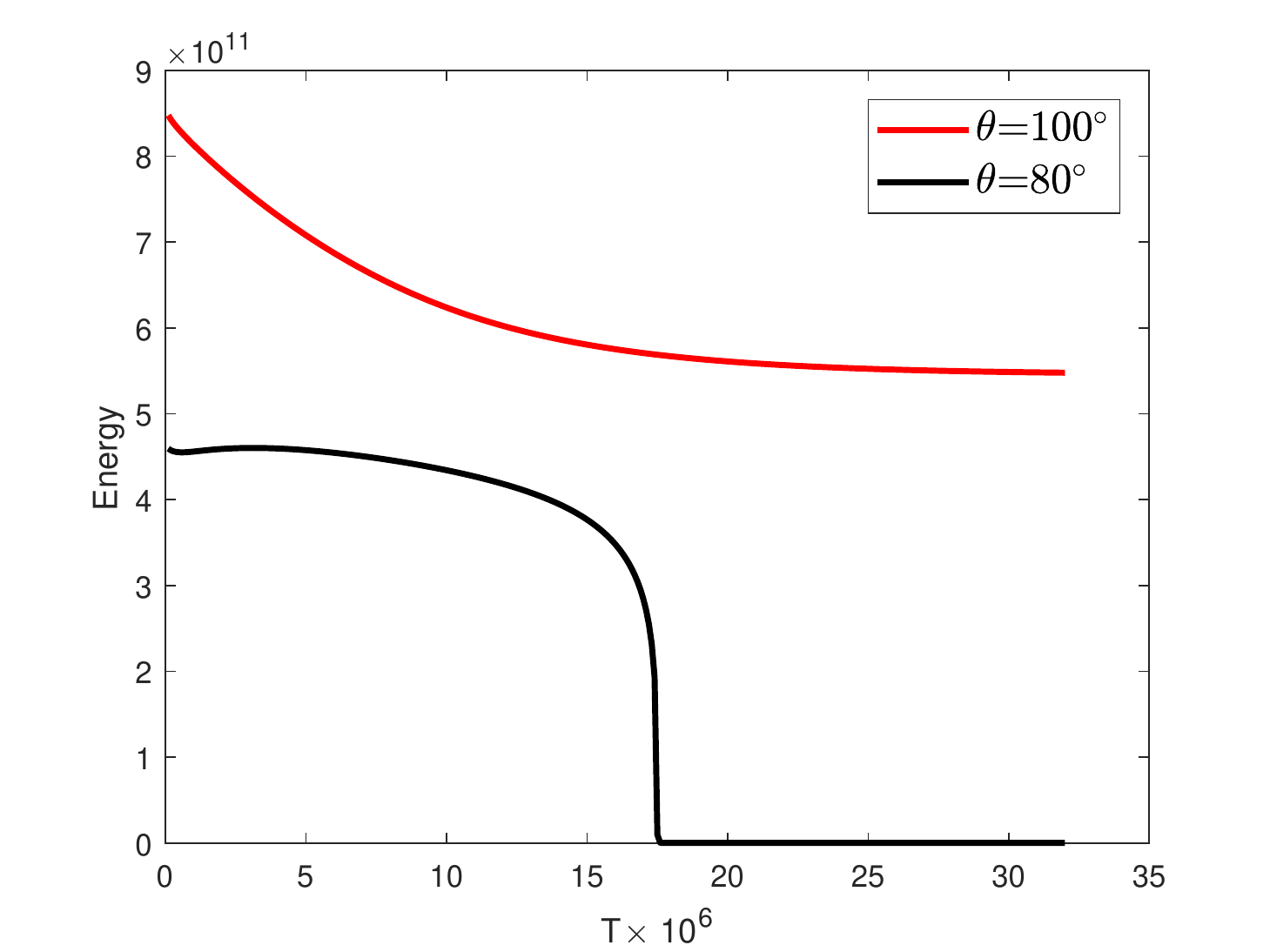}}
\caption{Time evolution of the system energy of 2D drops.}
\label{fig:2D_energy_vs_time}
\end{figure}

To further illustrate the effect of contact angle on critical radius, a drop of radius $R=33$ in contact with the solid surface with $\theta=80\degree$ or $100\degree$ is simulated in a domain size of $300\times 300$. According to Eq.~(\ref{eq:criticalRadius2D}), the critical radiuses for $\theta=80\degree$ and $100\degree$ are $36$ and $31$, respectively.
Thus, the drop with $\theta=80\degree$ will vanish while the drop with $\theta=100\degree$ will be maintained. The shrinkage processes for both contact angles are shown in Fig.~\ref{fig:2Dtheta80_100}. The contours of $\phi=0$ are plotted to indicate the interface locations. It can be observed that the drop with $\theta=80\degree$ gradually vanishes while the drop with $\theta=100\degree$ is close to a steady-state shape, which is in line with the theoretical prediction. As the Cahn-Hilliard dynamics tends to minimize the system energy, the equilibrium state of a drop should correspond to the minimum energy of the system. Thus, we also plot the energy curves of the above two drops, as shown in Fig.~\ref{fig:2D_energy_vs_time}. As seen, the energy of the drop with $\theta=80\degree$ finally decreases to zero while the total energy of the drop with $\theta=100\degree$ approaches a non-zero stable value, which
confirms the theoretical predictions Eq.~(\ref{eq:criticalRadius2D}) and Eq.(\ref{eq:critical3D}).

In this paper, we carried out a theoretical analysis of the spontaneous shrinkage of a drop on a flat surface with a wetting boundary condition. It is shown that there exists a critical radius for the drop  related to the domain size, interface width and the contact angle.
All drops below the critical radius will eventually vanish. The critical radius could be rather large for a small contact angle for a given  domain and interface width. Although drop shrinkage is very slow, it may be important for some practical applications with wetting phenomena, such as prediction of the relative permeability in porous media with the Cahn-Hilliard model .

We thank Pentau Yue for helpful discussions. This study was supported by the National Science Foundation of China(51836003).
\section*{References}
\bibliography{mybib}

 \newcommand{\noop}[1]{}
\begin{thebibliography}{20}%
\makeatletter
\providecommand \@ifxundefined [1]{%
 \@ifx{#1\undefined}
}%
\providecommand \@ifnum [1]{%
 \ifnum #1\expandafter \@firstoftwo
 \else \expandafter \@secondoftwo
 \fi
}%
\providecommand \@ifx [1]{%
 \ifx #1\expandafter \@firstoftwo
 \else \expandafter \@secondoftwo
 \fi
}%
\providecommand \natexlab [1]{#1}%
\providecommand \enquote  [1]{``#1''}%
\providecommand \bibnamefont  [1]{#1}%
\providecommand \bibfnamefont [1]{#1}%
\providecommand \citenamefont [1]{#1}%
\providecommand \href@noop [0]{\@secondoftwo}%
\providecommand \href [0]{\begingroup \@sanitize@url \@href}%
\providecommand \@href[1]{\@@startlink{#1}\@@href}%
\providecommand \@@href[1]{\endgroup#1\@@endlink}%
\providecommand \@sanitize@url [0]{\catcode `\\12\catcode `\$12\catcode
  `\&12\catcode `\#12\catcode `\^12\catcode `\_12\catcode `\%12\relax}%
\providecommand \@@startlink[1]{}%
\providecommand \@@endlink[0]{}%
\providecommand \url  [0]{\begingroup\@sanitize@url \@url }%
\providecommand \@url [1]{\endgroup\@href {#1}{\urlprefix }}%
\providecommand \urlprefix  [0]{URL }%
\providecommand \Eprint [0]{\href }%
\providecommand \doibase [0]{http://dx.doi.org/}%
\providecommand \selectlanguage [0]{\@gobble}%
\providecommand \bibinfo  [0]{\@secondoftwo}%
\providecommand \bibfield  [0]{\@secondoftwo}%
\providecommand \translation [1]{[#1]}%
\providecommand \BibitemOpen [0]{}%
\providecommand \bibitemStop [0]{}%
\providecommand \bibitemNoStop [0]{.\EOS\space}%
\providecommand \EOS [0]{\spacefactor3000\relax}%
\providecommand \BibitemShut  [1]{\csname bibitem#1\endcsname}%
\let\auto@bib@innerbib\@empty
\bibitem [{\citenamefont {Karma}(2001)}]{karma2001phase}%
  \BibitemOpen
  \bibfield  {author} {\bibinfo {author} {\bibfnamefont {A.}~\bibnamefont
  {Karma}},\ }\href@noop {} {\bibfield  {journal} {\bibinfo  {journal} {Phys.
  Rev. Lett}\ }\textbf {\bibinfo {volume} {87}},\ \bibinfo {pages} {115701}
  (\bibinfo {year} {2001})}\BibitemShut {NoStop}%
\bibitem [{\citenamefont {Rojas}\ \emph {et~al.}(2015)\citenamefont {Rojas},
  \citenamefont {Takaki},\ and\ \citenamefont {Ohno}}]{rojas2015phase}%
  \BibitemOpen
  \bibfield  {author} {\bibinfo {author} {\bibfnamefont {R.}~\bibnamefont
  {Rojas}}, \bibinfo {author} {\bibfnamefont {T.}~\bibnamefont {Takaki}}, \
  and\ \bibinfo {author} {\bibfnamefont {M.}~\bibnamefont {Ohno}},\ }\href@noop
  {} {\bibfield  {journal} {\bibinfo  {journal} {J. Comput. Phys.}\ }\textbf
  {\bibinfo {volume} {298}},\ \bibinfo {pages} {29} (\bibinfo {year}
  {2015})}\BibitemShut {NoStop}%
\bibitem [{\citenamefont {Nestler}\ \emph {et~al.}(2005)\citenamefont
  {Nestler}, \citenamefont {Garcke},\ and\ \citenamefont
  {Stinner}}]{nestler2005multicomponent}%
  \BibitemOpen
  \bibfield  {author} {\bibinfo {author} {\bibfnamefont {B.}~\bibnamefont
  {Nestler}}, \bibinfo {author} {\bibfnamefont {H.}~\bibnamefont {Garcke}}, \
  and\ \bibinfo {author} {\bibfnamefont {B.}~\bibnamefont {Stinner}},\
  }\href@noop {} {\bibfield  {journal} {\bibinfo  {journal} {Phys. Rev. E}\
  }\textbf {\bibinfo {volume} {71}},\ \bibinfo {pages} {041609} (\bibinfo
  {year} {2005})}\BibitemShut {NoStop}%
\bibitem [{\citenamefont {Ambati}\ \emph {et~al.}(2015)\citenamefont {Ambati},
  \citenamefont {Gerasimov},\ and\ \citenamefont
  {De~Lorenzis}}]{ambati2015review}%
  \BibitemOpen
  \bibfield  {author} {\bibinfo {author} {\bibfnamefont {M.}~\bibnamefont
  {Ambati}}, \bibinfo {author} {\bibfnamefont {T.}~\bibnamefont {Gerasimov}}, \
  and\ \bibinfo {author} {\bibfnamefont {L.}~\bibnamefont {De~Lorenzis}},\
  }\href@noop {} {\bibfield  {journal} {\bibinfo  {journal} {Comput Mech}\
  }\textbf {\bibinfo {volume} {55}},\ \bibinfo {pages} {383} (\bibinfo {year}
  {2015})}\BibitemShut {NoStop}%
\bibitem [{\citenamefont {Pons}\ and\ \citenamefont
  {Karma}(2010)}]{pons2010helical}%
  \BibitemOpen
  \bibfield  {author} {\bibinfo {author} {\bibfnamefont {A.~J.}\ \bibnamefont
  {Pons}}\ and\ \bibinfo {author} {\bibfnamefont {A.}~\bibnamefont {Karma}},\
  }\href@noop {} {\bibfield  {journal} {\bibinfo  {journal} {Nature}\ }\textbf
  {\bibinfo {volume} {464}},\ \bibinfo {pages} {85} (\bibinfo {year}
  {2010})}\BibitemShut {NoStop}%
\bibitem [{\citenamefont {Badalassi}\ \emph {et~al.}(2003)\citenamefont
  {Badalassi}, \citenamefont {Ceniceros},\ and\ \citenamefont
  {Banerjee}}]{badalassi2003computation}%
  \BibitemOpen
  \bibfield  {author} {\bibinfo {author} {\bibfnamefont {V.}~\bibnamefont
  {Badalassi}}, \bibinfo {author} {\bibfnamefont {H.}~\bibnamefont
  {Ceniceros}}, \ and\ \bibinfo {author} {\bibfnamefont {S.}~\bibnamefont
  {Banerjee}},\ }\href@noop {} {\bibfield  {journal} {\bibinfo  {journal} {J.
  Comput. Phys.}\ }\textbf {\bibinfo {volume} {190}},\ \bibinfo {pages} {371}
  (\bibinfo {year} {2003})}\BibitemShut {NoStop}%
\bibitem [{\citenamefont {Shen}\ and\ \citenamefont
  {Yang}(2015)}]{shen2015decoupled}%
  \BibitemOpen
  \bibfield  {author} {\bibinfo {author} {\bibfnamefont {J.}~\bibnamefont
  {Shen}}\ and\ \bibinfo {author} {\bibfnamefont {X.}~\bibnamefont {Yang}},\
  }\href@noop {} {\bibfield  {journal} {\bibinfo  {journal} {Siam. J. Numer.
  Anal.}\ }\textbf {\bibinfo {volume} {53}},\ \bibinfo {pages} {279} (\bibinfo
  {year} {2015})}\BibitemShut {NoStop}%
\bibitem [{\citenamefont {Ding}\ and\ \citenamefont
  {Spelt}(2007)}]{ding2007wetting}%
  \BibitemOpen
  \bibfield  {author} {\bibinfo {author} {\bibfnamefont {H.}~\bibnamefont
  {Ding}}\ and\ \bibinfo {author} {\bibfnamefont {P.~D.}\ \bibnamefont
  {Spelt}},\ }\href@noop {} {\bibfield  {journal} {\bibinfo  {journal} {Phys.
  Rev. E}\ }\textbf {\bibinfo {volume} {75}},\ \bibinfo {pages} {046708}
  (\bibinfo {year} {2007})}\BibitemShut {NoStop}%
\bibitem [{\citenamefont {Yue}\ and\ \citenamefont {Feng}(2011)}]{yue2011can}%
  \BibitemOpen
  \bibfield  {author} {\bibinfo {author} {\bibfnamefont {P.}~\bibnamefont
  {Yue}}\ and\ \bibinfo {author} {\bibfnamefont {J.}~\bibnamefont {Feng}},\
  }\href@noop {} {\bibfield  {journal} {\bibinfo  {journal} {Eur. Phys. J Spec.
  Top}\ }\textbf {\bibinfo {volume} {197}},\ \bibinfo {pages} {37} (\bibinfo
  {year} {2011})}\BibitemShut {NoStop}%
\bibitem [{\citenamefont {Teigen}\ \emph {et~al.}(2011)\citenamefont {Teigen},
  \citenamefont {Song}, \citenamefont {Lowengrub},\ and\ \citenamefont
  {Voigt}}]{teigen2011diffuse}%
  \BibitemOpen
  \bibfield  {author} {\bibinfo {author} {\bibfnamefont {K.~E.}\ \bibnamefont
  {Teigen}}, \bibinfo {author} {\bibfnamefont {P.}~\bibnamefont {Song}},
  \bibinfo {author} {\bibfnamefont {J.}~\bibnamefont {Lowengrub}}, \ and\
  \bibinfo {author} {\bibfnamefont {A.}~\bibnamefont {Voigt}},\ }\href@noop {}
  {\bibfield  {journal} {\bibinfo  {journal} {J. Comput. Phys.}\ }\textbf
  {\bibinfo {volume} {230}},\ \bibinfo {pages} {375} (\bibinfo {year}
  {2011})}\BibitemShut {NoStop}%
\bibitem [{\citenamefont {Liu}\ and\ \citenamefont
  {Shen}(2003)}]{liu2003phase}%
  \BibitemOpen
  \bibfield  {author} {\bibinfo {author} {\bibfnamefont {C.}~\bibnamefont
  {Liu}}\ and\ \bibinfo {author} {\bibfnamefont {J.}~\bibnamefont {Shen}},\
  }\href@noop {} {\bibfield  {journal} {\bibinfo  {journal} {Physica D.}\
  }\textbf {\bibinfo {volume} {179}},\ \bibinfo {pages} {211} (\bibinfo {year}
  {2003})}\BibitemShut {NoStop}%
\bibitem [{\citenamefont {Yue}\ \emph {et~al.}(2004)\citenamefont {Yue},
  \citenamefont {Feng}, \citenamefont {Liu},\ and\ \citenamefont
  {Shen}}]{yue2004diffuse}%
  \BibitemOpen
  \bibfield  {author} {\bibinfo {author} {\bibfnamefont {P.}~\bibnamefont
  {Yue}}, \bibinfo {author} {\bibfnamefont {J.~J.}\ \bibnamefont {Feng}},
  \bibinfo {author} {\bibfnamefont {C.}~\bibnamefont {Liu}}, \ and\ \bibinfo
  {author} {\bibfnamefont {J.}~\bibnamefont {Shen}},\ }\href@noop {} {\bibfield
   {journal} {\bibinfo  {journal} {J. Fluid Mech}\ }\textbf {\bibinfo {volume}
  {515}},\ \bibinfo {pages} {293} (\bibinfo {year} {2004})}\BibitemShut
  {NoStop}%
\bibitem [{\citenamefont {Lee}\ and\ \citenamefont
  {Fischer}(2006)}]{lee2006eliminating}%
  \BibitemOpen
  \bibfield  {author} {\bibinfo {author} {\bibfnamefont {T.}~\bibnamefont
  {Lee}}\ and\ \bibinfo {author} {\bibfnamefont {P.~F.}\ \bibnamefont
  {Fischer}},\ }\href@noop {} {\bibfield  {journal} {\bibinfo  {journal} {Phys.
  Rev. E}\ }\textbf {\bibinfo {volume} {74}},\ \bibinfo {pages} {046709}
  (\bibinfo {year} {2006})}\BibitemShut {NoStop}%
\bibitem [{\citenamefont {Yue}\ \emph {et~al.}(2007)\citenamefont {Yue},
  \citenamefont {Zhou},\ and\ \citenamefont {Feng}}]{yue2007spontaneous}%
  \BibitemOpen
  \bibfield  {author} {\bibinfo {author} {\bibfnamefont {P.}~\bibnamefont
  {Yue}}, \bibinfo {author} {\bibfnamefont {C.}~\bibnamefont {Zhou}}, \ and\
  \bibinfo {author} {\bibfnamefont {J.~J.}\ \bibnamefont {Feng}},\ }\href@noop
  {} {\bibfield  {journal} {\bibinfo  {journal} {J. Comput. Phys.}\ }\textbf
  {\bibinfo {volume} {223}},\ \bibinfo {pages} {1} (\bibinfo {year}
  {2007})}\BibitemShut {NoStop}%
\bibitem [{\citenamefont {Zheng}\ \emph {et~al.}(2014)\citenamefont {Zheng},
  \citenamefont {Lee}, \citenamefont {Guo},\ and\ \citenamefont
  {Rumschitzki}}]{zheng2014shrinkage}%
  \BibitemOpen
  \bibfield  {author} {\bibinfo {author} {\bibfnamefont {L.}~\bibnamefont
  {Zheng}}, \bibinfo {author} {\bibfnamefont {T.}~\bibnamefont {Lee}}, \bibinfo
  {author} {\bibfnamefont {Z.}~\bibnamefont {Guo}}, \ and\ \bibinfo {author}
  {\bibfnamefont {D.}~\bibnamefont {Rumschitzki}},\ }\href@noop {} {\bibfield
  {journal} {\bibinfo  {journal} {Phys. Rev. E}\ }\textbf {\bibinfo {volume}
  {89}},\ \bibinfo {pages} {033302} (\bibinfo {year} {2014})}\BibitemShut
  {NoStop}%
\bibitem [{\citenamefont {Bonn}\ \emph {et~al.}(2009)\citenamefont {Bonn},
  \citenamefont {Eggers}, \citenamefont {Indekeu}, \citenamefont {Meunier},\
  and\ \citenamefont {Rolley}}]{bonn2009wetting}%
  \BibitemOpen
  \bibfield  {author} {\bibinfo {author} {\bibfnamefont {D.}~\bibnamefont
  {Bonn}}, \bibinfo {author} {\bibfnamefont {J.}~\bibnamefont {Eggers}},
  \bibinfo {author} {\bibfnamefont {J.}~\bibnamefont {Indekeu}}, \bibinfo
  {author} {\bibfnamefont {J.}~\bibnamefont {Meunier}}, \ and\ \bibinfo
  {author} {\bibfnamefont {E.}~\bibnamefont {Rolley}},\ }\href@noop {}
  {\bibfield  {journal} {\bibinfo  {journal} {Rev. Mod. Phys.}\ }\textbf
  {\bibinfo {volume} {81}},\ \bibinfo {pages} {739} (\bibinfo {year}
  {2009})}\BibitemShut {NoStop}%
\bibitem [{\citenamefont {Sohrabi}\ \emph {et~al.}(2011)\citenamefont
  {Sohrabi}, \citenamefont {Kechut}, \citenamefont {Riazi}, \citenamefont
  {Jamiolahmady}, \citenamefont {Ireland},\ and\ \citenamefont
  {Robertson}}]{sohrabi2011safe}%
  \BibitemOpen
  \bibfield  {author} {\bibinfo {author} {\bibfnamefont {M.}~\bibnamefont
  {Sohrabi}}, \bibinfo {author} {\bibfnamefont {N.~I.}\ \bibnamefont {Kechut}},
  \bibinfo {author} {\bibfnamefont {M.}~\bibnamefont {Riazi}}, \bibinfo
  {author} {\bibfnamefont {M.}~\bibnamefont {Jamiolahmady}}, \bibinfo {author}
  {\bibfnamefont {S.}~\bibnamefont {Ireland}}, \ and\ \bibinfo {author}
  {\bibfnamefont {G.}~\bibnamefont {Robertson}},\ }\href@noop {} {\bibfield
  {journal} {\bibinfo  {journal} {Chem. Eng. Res. Des}\ }\textbf {\bibinfo
  {volume} {89}},\ \bibinfo {pages} {1865} (\bibinfo {year}
  {2011})}\BibitemShut {NoStop}%
\bibitem [{\citenamefont {Cahn}\ and\ \citenamefont
  {Hilliard}(1958)}]{cahn1958free}%
  \BibitemOpen
  \bibfield  {author} {\bibinfo {author} {\bibfnamefont {J.~W.}\ \bibnamefont
  {Cahn}}\ and\ \bibinfo {author} {\bibfnamefont {J.~E.}\ \bibnamefont
  {Hilliard}},\ }\href@noop {} {\bibfield  {journal} {\bibinfo  {journal} {J.
  Chem. Phys.}\ }\textbf {\bibinfo {volume} {28}},\ \bibinfo {pages} {258}
  (\bibinfo {year} {1958})}\BibitemShut {NoStop}%
\bibitem [{\citenamefont {Yue}\ \emph {et~al.}(2010)\citenamefont {Yue},
  \citenamefont {Zhou},\ and\ \citenamefont {Feng}}]{yue2010sharp}%
  \BibitemOpen
  \bibfield  {author} {\bibinfo {author} {\bibfnamefont {P.}~\bibnamefont
  {Yue}}, \bibinfo {author} {\bibfnamefont {C.}~\bibnamefont {Zhou}}, \ and\
  \bibinfo {author} {\bibfnamefont {J.~J.}\ \bibnamefont {Feng}},\ }\href@noop
  {} {\bibfield  {journal} {\bibinfo  {journal} {J. Fluid Mech}\ }\textbf
  {\bibinfo {volume} {645}},\ \bibinfo {pages} {279} (\bibinfo {year}
  {2010})}\BibitemShut {NoStop}%
\bibitem [{\citenamefont {Zhang}\ \emph {et~al.}(2019)\citenamefont {Zhang},
  \citenamefont {Guo},\ and\ \citenamefont {Liang}}]{zhang2019high}%
  \BibitemOpen
  \bibfield  {author} {\bibinfo {author} {\bibfnamefont {C.}~\bibnamefont
  {Zhang}}, \bibinfo {author} {\bibfnamefont {Z.}~\bibnamefont {Guo}}, \ and\
  \bibinfo {author} {\bibfnamefont {H.}~\bibnamefont {Liang}},\ }\href@noop {}
  {\bibfield  {journal} {\bibinfo  {journal} {Phys. Rev. E}\ }\textbf {\bibinfo
  {volume} {99}},\ \bibinfo {pages} {043310} (\bibinfo {year}
  {2019})}\BibitemShut {NoStop}%
\end{thebibliography}%
\end{document}